\documentclass{article}
\usepackage{ismir,amsmath,cite,url}
\usepackage{graphicx}
\usepackage{color}
\usepackage{tikz}
\usepackage{graphicx}
\usepackage[nomessages]{fp}
\usepackage{amssymb,mathtools}
\usepackage{epstopdf}

\title{The ACCompanion v0.1:  
An Expressive Accompaniment System}

\multauthor
{Carlos Cancino-Chac\'on$^{1,2}$ \hspace{1cm} Martin Bonev$^{3,1}$ \hspace{1cm}Amaury Durand$^{2}$} { \bfseries{Maarten Grachten$^2$ \hspace{1cm} Andreas Arzt $^2$ \hspace{1cm} Laura Bishop$^1$} \\ 
\bfseries{Werner Goebl$^3$ \hspace{1cm} Gerhard Widmer$^{1,2}$}\\
  $^1$ Austrian Research Institute for Artificial Intelligence, Vienna, Austria\\
$^2$ Department of Computational Perception, Johannes Kepler University Linz, Austria\\
$^3$  Department of Music Acoustics, University of Music and Performing Arts Vienna, Austria\\
{\tt\small carlos.cancino@ofai.at }
}
\def\authorname{Cancino-Chac\'on, Bonev, Durand, Grachten, Arzt, Bishop, Goebl, Widmer}

\begin{document}

\maketitle
\begin{abstract}
In this paper we present a preliminary version of the ACCompanion, an expressive accompaniment system for MIDI input.
The system uses a probabilistic monophonic score follower to track the position of the soloist in the score, and a linear Gaussian model to compute tempo updates.
The expressiveness of the system is powered by the Basis-Mixer, a state-of-the-art computational model of expressive music performance.
The system allows for expressive dynamics, timing and articulation.

\end{abstract}
\section{Introduction}\label{sec:introduction}

Computational accompaniment systems attempt to automatically generate synchronized accompaniment for a (human) solo performance, usually in real time.
While most of the work on accompaniment systems has focused on the problem of score following  \cite{Cont:2012im, Nakamura:2015to,Raphael:2010Bs}, 
in recent years there has been increased interest in exploring expressive accompaniment systems. 
For example, Xia et al.~\cite{Xia:2015ur} present an accompaniment system capable of generating expressive dynamics and timing using linear dynamical systems.

In this demo, we present a preliminary version of an accompaniment system, the ACCompanion. 
This system combines a monophonic score follower based on Hidden Markov Models (HMMs) with the Basis-Mixer (BM)~\cite{CancinoChacon:2017ht}, a neural network-powered framework for expressive music performance.


\section{Score Following}\label{sec:score_following}
The ACCompanion uses an HMM-based monophonic score follower.
The observed variables in this HMM model are the performed MIDI pitches and the inter-onset-intervals (IOIs), i.e., the time intervals between consecutive notes in the solo part; the hidden variable is the position in the score.
We use a switching Kalman filter which considers the beat period to be a linear Gaussian process in which the transition matrices and noise parameters depend on the hidden state of the HMM~\cite{Murphy:1998wd}.

\section{The Basis-Mixer (BM): \\ an expressive performer}\label{sec:basis_mixer}
\begin{figure}[t]
\begin{center}
\textbf{Onsetwise Model} \\

\begin{tikzpicture}[->, thick, bend angle=35]
\tiny
\newcommand{\ovscaling}{0.75}
\newcommand{\hscale}{\ovscaling * 1.5}
\tikzstyle{main}=[circle, minimum size = \ovscaling * 2.5mm, thick, draw =black!80]
\foreach \name in {1,...,3}
    \node[main, fill = gray!10] (l\name) at (\hscale*\name, \ovscaling * 2.75) {$y_{o_\name}$};
\foreach \name in {1,...,3}
    \node[main, fill = gray!40] (i\name) at (\hscale*\name,\ovscaling * 1.25) {$\mathbf{h}_{\name; \mathrm{bw}}$};
\foreach \name in {1,...,3}
    \node[main, fill = gray!40] (h\name) at (\hscale*\name,\ovscaling * -0.25) {$\mathbf{h}_{\name; \mathrm{fw}}$};
\foreach \name in {1,...,3}
    \node[main, fill = white!100] (x\name) at (\hscale*\name,\ovscaling * -2) {$\boldsymbol\varphi(o_\name)$};
\foreach \h in {1,...,3}
       {
        \path (x\h) edge [bend right] (i\h);
        \path (x\h) edge (h\h);
        \path (h\h) edge [bend left] (l\h);
        \path (i\h) edge (l\h);
       }
       \foreach \name in {5,...,7}
       \FPeval{\nodename}{clip(\name + 1)}
    \node[main, fill = gray!10] (l\name) at (\hscale*\name,2.75* \ovscaling) {$y_{o_{\nodename}}$};
\foreach \name in {5,...,7}
	\FPeval{\nodename}{clip(\name + 1)}
    \node[main, fill = gray!40] (i\name) at (\hscale*\name,1.25*\ovscaling) {$\mathbf{h}_{\nodename; \mathrm{bw}}$};
\foreach \name in {5,...,7}
	\FPeval{\nodename}{clip(\name + 1)}
    \node[main, fill = gray!40] (h\name) at (\hscale*\name,-0.25*\ovscaling) {$\mathbf{h}_{\nodename; \mathrm{fw}}$};
\foreach \name in {5,...,7}
	\FPeval{\nodename}{clip(\name + 1)}
    \node[main, fill = white!100] (x\name) at (\hscale*\name,-2.0*\ovscaling) {$\boldsymbol\varphi(o_{\nodename})$};
\foreach \h in {5,...,7}
       {
        \path (x\h) edge [bend right] (i\h);
        \path (x\h) edge (h\h);
        \path (h\h) edge [bend left] (l\h);
        \path (i\h) edge (l\h);
       }
       \node (x4) at (\hscale*4, -2*\ovscaling) {$\dots$};
       \node (h4) at (\hscale*4, -0.25*\ovscaling) {$\dots$};
       \node (i4) at (\hscale*4, 1.25*\ovscaling) {$\dots$};
       \node (l4) at (\hscale*4, 2.75*\ovscaling) {$\dots$};

\foreach \current/\next in {1/2,2/3,3/4,4/5,5/6,6/7} 
       {
        \path (i\next) edge (i\current);
        \path (h\current) edge (h\next);
       }
\end{tikzpicture}

\textbf{Notewise Model}\\

\begin{tikzpicture}[->,thick]
\tiny
\newcommand{\ovscaling}{0.75}
\newcommand{\hscale}{\ovscaling * 1.5}
\tikzstyle{main}=[circle, minimum size = \ovscaling * 2.5mm, thick, draw =black!80]
\foreach \name in {1,...,3}
    \node[main, fill = gray!10] (l\name) at (\hscale*\name, \ovscaling * 2.75) {$y_{n_\name}$};
\foreach \name in {1,...,3}
    \node[main, fill = gray!40] (i\name) at (\hscale*\name, \ovscaling * 1.25) {$\mathbf{h}_{\name}^{(2)}$};
\foreach \name in {1,...,3}
    \node[main, fill = gray!40] (h\name) at (\hscale*\name, \ovscaling * -0.25) {$\mathbf{h}_{\name}^{(1)}$};
\foreach \name in {1,...,3}
    \node[main, fill = white!100] (x\name) at (\hscale*\name, \ovscaling * -2) {$\boldsymbol\varphi(n_{\name})$};
\foreach \h in {1,...,3}
       {
        \path (x\h) edge (h\h);
        \path (h\h) edge (i\h);
        \path (i\h) edge (l\h);
       }
       \node (dotsx) at (\hscale*4, -2*\ovscaling) {$\dots$};
       \node (dotsh) at (\hscale*4, -0.25*\ovscaling) {$\dots$};
       \node (dotsi) at (\hscale*4, 1.25*\ovscaling) {$\dots$};
       \node (dotsl) at (\hscale*4, 2.75*\ovscaling) {$\dots$};
       \foreach \name in {5,...,7}
       \FPeval{\nodename}{clip(\name + 1)}
    \node[main, fill = gray!10] (l\name) at (\hscale*\name, \ovscaling * 2.75) {$y_{n_{\nodename}}$};
\foreach \name in {5,...,7}
\FPeval{\nodename}{clip(\name + 1)}
    \node[main, fill = gray!40] (i\name) at (\hscale*\name, \ovscaling * 1.25) {$\mathbf{h}_{\nodename}^{(2)}$};
\foreach \name in {5,...,7}
\FPeval{\nodename}{clip(\name + 1)}
    \node[main, fill = gray!40] (h\name) at (\hscale*\name, \ovscaling * -0.25) {$\mathbf{h}_{\nodename}^{(1)}$};
\foreach \name in {5,...,7}
\FPeval{\nodename}{clip(\name + 1)}
    \node[main, fill = white!100] (x\name) at (\hscale*\name, \ovscaling * -2) {$\boldsymbol\varphi(n_{\nodename})$};
\foreach \h in {5,...,7}
       {
        \path (x\h) edge (h\h);
        \path (h\h) edge (i\h);
        \path (i\h) edge (l\h);
       }
\end{tikzpicture}
\caption{Neural architectures used for the BM models.}
\label{fig:model_architectures}
\end{center}
\end{figure}
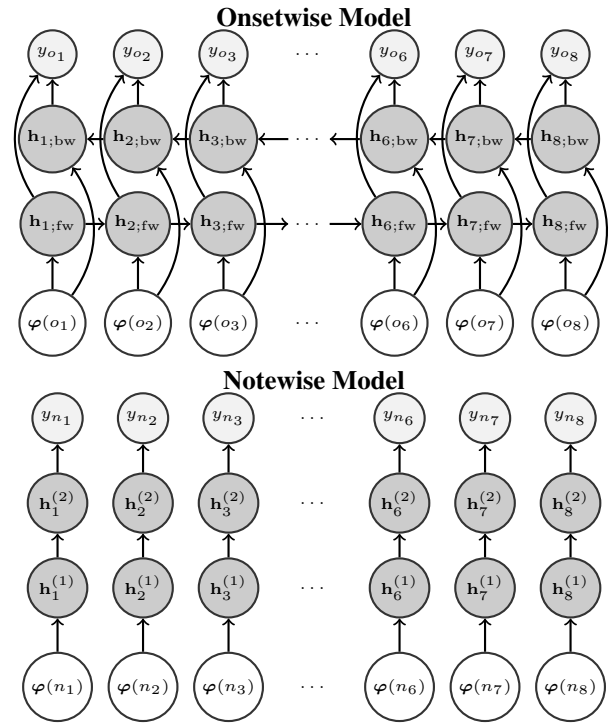
The BM framework encodes a musical score using \emph{basis functions}, i.e., numerical descriptors that represent certain structural aspects of the score.
Using this framework, the performance of the accompaniment part of a piece is encoded in five target variables, to be predicted for every note in the accompaniment:
\begin{enumerate}
\item $\mathit{Loudness}_{\mathit{trend}}$: Ratio of the maximal MIDI velocity at each accompaniment onset to the corresponding MIDI velocity of the solo performance.
\item $\mathit{BP}$: Ratio of the beat period at each accompaniment onset to the corresponding beat period of the solo performance, as estimated by the score follower.
\item $\mathit{Loudness}_{\mathit{dev}}$: Deviation of each accompaniment note velocity from $\mathit{Loudness}_{\mathit{trend}}$  at each onset.
\item $\mathit{Timing}$: Micro-deviations of each note in a chord from the average onset time in the accompaniment.
\item $\mathit{Articulation}$ : Ratio of the performed IOIs in seconds to the score IOIs in beats.
\end{enumerate}
The BM uses bidirectional recurrent neural networks to predict the onsetwise targets and feed forward neural networks to predict the notewise targets.
These networks are illustrated in Figure \ref{fig:model_architectures}.
The BM is trained using a dataset of performances of Beethoven Piano Sonatas~\cite{CancinoChacon:2017ht}.

\section{The ACCompanion}\label{sec:accompanion}
\begin{figure}[t]
\begin{center}
\includegraphics[width=0.95\linewidth]{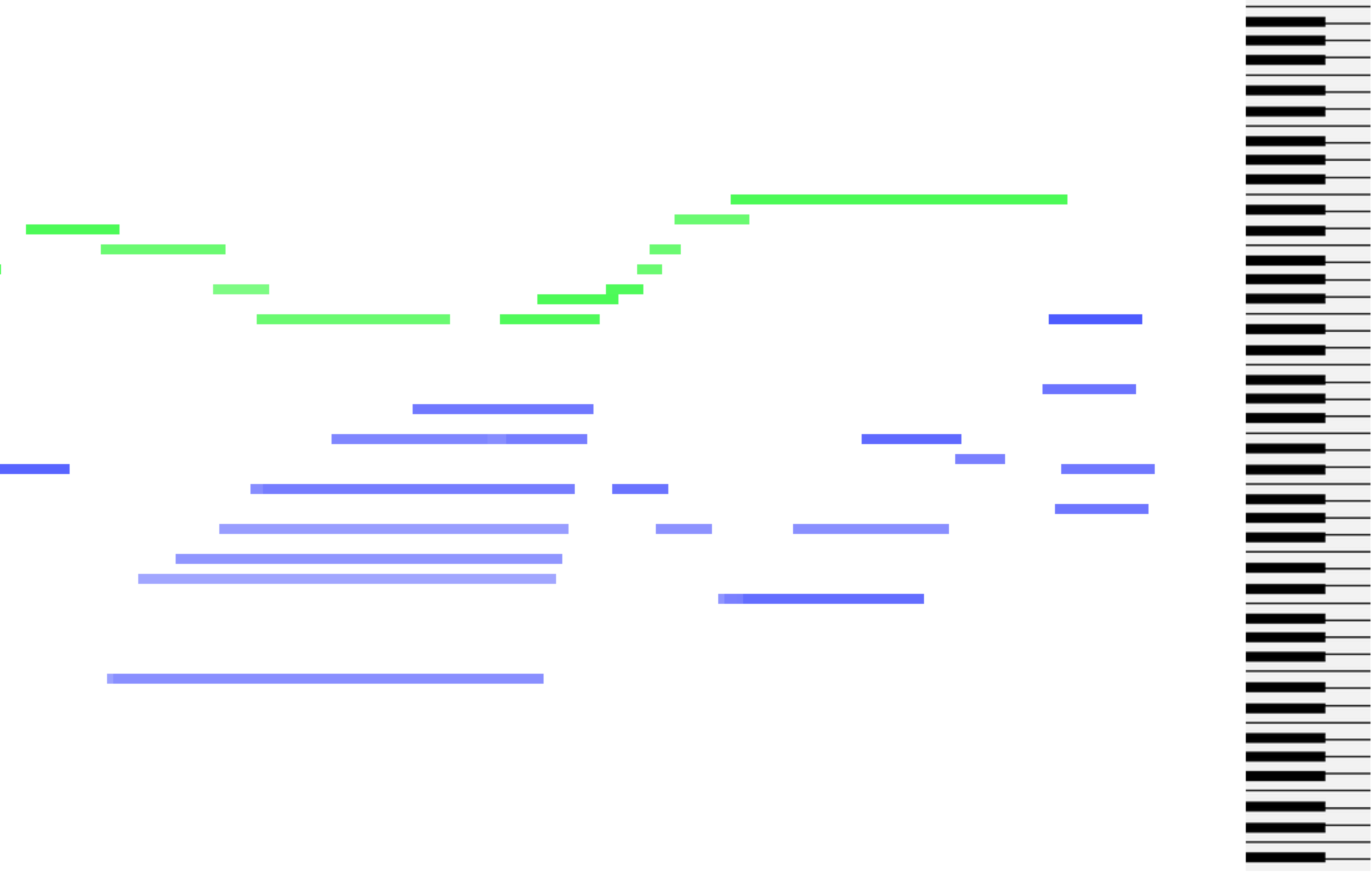}
\caption{Screenshot of the ACCompanion visualization.}\label{fig:gui}
\end{center}
\end{figure}

The system is implemented using Python.
The performance of a piece is presented visually by a Graphical User Interface (GUI), which displays both the solo and accompaniment parts in a piano roll in real time. 
A simple color scheme is used to illustrate the loudness, with brighter color lines representing louder notes. 
Solo and accompaniment parts are distinguished using different colors. 
In the solo part, green lines represent correctly played notes, while inserted and misplayed notes are drawn in red. 
The accompaniment part is presented with blue lines.
A screenshot of the GUI is shown in Figure \ref{fig:gui}.
For the demo, we use a USB knob controller, a PowerMate by Griffin Technology\footnote{\url{https://griffintechnology.com/us/products/stylus-keyboards/powermate}}, to exaggerate or minimize certain aspects of the expressive performance by controlling the scaling of the expressive targets in real time.

\section{Conclusions}\label{sec:conclusions}
We have presented the first simple prototype of an automatic accompaniment system that has its own
ideas (model) of expressive performance.
Future versions of the ACCompanion will include more sophisticated polyphonic score followers, following the work by \cite{Nakamura:2015to}.
Also, we aim at integrating more complex variants of the BM framework for expressive performance, 
trained on real ensemble performance data.

For this early prototype, we have not yet done any thorough evaluation, other than playing with it in order to obtain a general impression. Later versions will, of course, be evaluated in much more systematic ways, for instance by measuring how well their performances correlate with those of human musicians.
However, this quantitative approach to evaluation is intrinsically problematic, as it makes very limiting assumptions regarding what kinds of performances are musically meaningful and `good'. Ultimately, really meaningful tests will
have to involve the judgment of human musicians and listeners. We defer this to later stages of the project, where we hope to have a more complete and musically sophisticated system.

\section{Acknowledgements}
This work has been funded by the European Research Council (ERC) under the EU’s Horizon 2020 Framework Programme (ERC Grant Agreement No. 670035, project CON ESPRESSIONE) and by the Austrian Science Fund (FWF) grant P29427.

\bibliography{bib/bib_cc}
\end{document}